\documentclass[aps,twocolumn]{revtex4}

\usepackage{graphicx}
\usepackage{amssymb}
\usepackage{dcolumn}
\usepackage{amsmath}
\usepackage{color}
\usepackage{hyperref}

\providecommand{\st}[1]{_{\text{#1}}}
\providecommand{\ut}[1]{^{\text{#1}}}
\newcommand{\bv}[1]{\mathbf{#1}}

\def\ueq{\ut{eq}}

\def\onehalf{\frac{1}{2}}

\def\uv{\bv{u}}

\def\Fv{\bv{F}}
\def\cv{\bv{c}}

\def\cv{\bv{c}}
\def\rv{\bv{r}}
\def\rsig{r_\sigma}
\def\RE{\mathrm{Re}}

\newcommand{\bitem}{\begin{itemize}}
\newcommand{\eitem}{\end{itemize}}
\newcommand{\benum}{\begin{enumerate}}
\newcommand{\eenum}{\end{enumerate}}
\newcommand{\btab}[1]{\begin{tabular}{#1}}
\newcommand{\etab}{\end{tabular}}
\newcommand{\btabn}[1]{\begin{tabular}{#1}}
\newcommand{\etabn}{\end{tabular}}
\newcommand{\beq}{\begin{equation}}
\newcommand{\eeq}{\end{equation}}
\newcommand{\beqn}{\begin{equation*}}
\newcommand{\eeqn}{\end{equation*}}
\newcommand{\bsplit}{\begin{split}}
\newcommand{\esplit}{\end{split}}

\begin{document}

\title{Viscous coalescence of droplets: a Lattice Boltzmann study}
\author{M. Gross}
\affiliation{Interdisciplinary Centre for Advanced Materials Simulation (ICAMS), Ruhr-Universit\"at Bochum, Universitaetsstr.\ 90a, 44789 Bochum, Germany}
\author{D. Raabe}
\affiliation{Max-Planck Institut f\"ur Eisenforschung, Max-Planck Str.~1, 40237 D\"usseldorf, Germany}
\author{I. Steinbach}
\affiliation{Interdisciplinary Centre for Advanced Materials Simulation (ICAMS), Ruhr-Universit\"at Bochum, Universitaetsstr.\ 90a, 44789 Bochum, Germany}
\author{F. Varnik}
\email{fathollah.varnik@rub.de}
\affiliation{Interdisciplinary Centre for Advanced Materials Simulation (ICAMS), Ruhr-Universit\"at Bochum, Universitaetsstr.\ 90a, 44789 Bochum, Germany}
\affiliation{Max-Planck Institut f\"ur Eisenforschung, Max-Planck Str.~1, 40237 D\"usseldorf, Germany}

\begin{abstract}
The coalescence of two resting liquid droplets in a saturated vapor phase is investigated by Lattice Boltzmann simulations in two and three dimensions. We find that, in the viscous regime, the bridge radius obeys a $t^{1/2}$-scaling law in time with the characteristic time scale given by the viscous time. 
Our results differ significantly from the predictions of existing analytical theories of viscous coalescence as well as from experimental observations. 
While the underlying reason for these deviations is presently unknown, a simple scaling argument is given that describes our results well.
\end{abstract}


\maketitle

\section{Introduction}
Coalescence of liquid droplets is important in many technological and natural phenomena, as, for example, coating and sintering processes \cite{solgel-science}, phase separation of emulsions \cite{nikolayev_prl1996}, or the formation of rain drops in clouds \cite{raindrops_1995}.
Coalescence is initiated when two droplets come into contact and form a liquid bridge, which then starts to grow due to surface tension.
This growth is typically either opposed by viscous dissipation or inertial forces, until finally the two droplets have merged to a single droplet.

\begin{figure}[b]
    \includegraphics[width=0.9\linewidth]{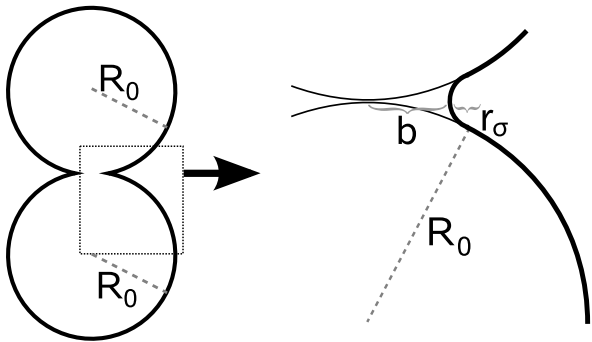}
   \caption{Sketch of two coalescing droplets. $R_0$ is the droplet radius, $b$ the radius of the connecting bridge and $r_\sigma$ the radius of curvature of the meniscus. The right image is a magnification of the bridge region.}
    \label{fig:droplets}
\end{figure}
Assuming that the initial growth of the liquid bridge just results from a competition between surface tension $\sigma$ and fluid viscosity $\eta$, the characteristic velocity scale is given by the capillary velocity, $u_c = \sigma/\eta$.
Taking the relevant length scale to be the size $b$ of the liquid bridge connecting the two droplets (see Fig.~\ref{fig:droplets}), a Reynolds number can be defined as
$\RE = \rho u_c b/\eta = \rho \sigma b/\eta^2$. Due to the presence of $b$, this number will always be small in the beginning of the process, regardless of the value of $\sigma$ and $\eta$.
The domain $\RE \lesssim 1$ then defines the \emph{viscous regime}, which can be described by the Stokes equations, in contrast to the \emph{inertial regime}, $\RE \gtrsim 1$, where the Euler equations hold.

Coalescence in the inertial regime is quite well understood nowadays
\cite{duchemin_coalescence_2003, inertial_coal, thoroddsen_2005, aarts_coalescence_2005, burton_prl2007, lee_eliminating_2006, paulsen_prl2011} and it has been established theoretically and experimentally that the evolution of the bridge radius in the inertial stage of coalescence follows a scaling law of the form $b\propto (R_0 \sigma / \rho)^{1/4} t^{1/2}$, where $R_0$ is the initial radius of each drop.

For the viscous regime, however, there remain some discrepancies between experiments and theory. The first effort to describe viscous coalescence dates back to the work of Frenkel \cite{frenkel_1949} on viscous sintering of solid particles.
By considering the balance between reduction of surface free energy and viscous dissipation, he derived a simple scaling law for the growth of the bridge radius,
$ b \propto \left(R_0\sigma/\eta \right)^{1/2} t^{1/2}$,
which was confirmed in subsequent sintering experiments \cite{sintering_exp}.
More sophisticated analysis of liquid droplet coalescence in the Stokes regime by Hopper \cite{hopper_1984} and Eggers et al.\ \cite{eggers_coalescence_1999}, aiming to overcome some of the simplifications of Frenkel's model, arrived at a characteristic $t^{0.86}$- or $t\log t$-growth law for the bridge radius at early times in the two- and three-dimensional case, respectively.
However, subsequent experiments \cite{aarts_coalescence_2005, yao_viscous_2005, burton_prl2007, paulsen_prl2011} neither reproduced Frenkel's nor Eggers' results, but instead reported a linear time dependence, $b\propto \sigma t/\eta$ in the viscous regime \cite{thoroddsen_remark}.
The disagreement between theory and experiments was usually attributed to the asymptotic nature of the analytical results, which are expected to hold only in the limit of very small bridge radii.
In a recent work, Paulsen et al.\ \cite{paulsen_pnas2012} address these discrepancies and argue that most experiments have in fact not operated in the actual Stokes regime, but instead in a different, ``inertially-limited'' viscous regime where the drops are governed by a balance between surface tension, viscous forces and the center-of-mass inertia. 

In this work, we study viscous coalescence of two droplets by Lattice Boltzmann computer simulations of a single-component, isothermal, non-ideal fluid in two and three dimensions. We consider two initially quiescent liquid droplets in a surrounding saturated vapor phase for various values of surface tension, viscosity and vapor density.
Interestingly, neither a growth of the bridge radius that is linear in time, as reported in experiments \cite{aarts_coalescence_2005, yao_viscous_2005, burton_prl2007, paulsen_prl2011}, nor a growth as predicted by theories of Stokesian coalescence \cite{hopper_1984, eggers_coalescence_1999} (i.e., $b\sim t^{0.86}$ in 2D or $b\sim -t \log t$ in 3D) is observed by us.
Instead, we find that the bridge radius is governed by an approximate $t^{1/2}$-scaling law, with the characteristic scale factor set by the viscous time -- similar to Frenkel's original result.
While we generally find evaporation and condensation processes taking place in our system, these are strongly suppressed compared to advection and do not seem to have a significant impact on the time evolution of the bridge radius.
Based on these findings, we provide a simple scaling argument that describes our results well.

%-------------------------------------------------------------------------------------------------------------

\section{Simulation model}

\subsection{Model}
Numerically, we solve the Navier-Stokes equations for a non-ideal fluid in both two and three dimensions with the Lattice Boltzmann (LB) method, which is a well-established tool for fluid dynamical simulations ranging from the micro- to the macroscale \cite{lbm_review}. Motivated by the growing interest in microfluidic applications, the LB method has been widely applied to the study of problems encompassing fluids in complex geometries or with multiple phases, such as droplet spreading or wetting \cite{droplets_gen}.
In the present work, we employ an implementation of the LB model described in \cite{lee_eliminating_2006}, which features faithful representation of isothermal two-phase thermo- and hydrodynamics \cite{kikkinides_2008, gross_shearstress_pre2011}. 

The LB method is based on an evolution equation for the distribution function $f(\rv,\cv)$ representing the probability to find a fluid element at a given position $\rv$ moving with a certain velocity $\cv_i$,
\beq
f_i(\rv+\cv_i, t+1) = f_i(\rv,t) - \frac{1}{\tau}\left[f_i(\rv,t)-f_i\ueq(\rv,t)\right] + f_i^F(\rv,t)\,.
\label{lbe}
\eeq
Here, $f_i(\rv)\equiv f(\rv,\cv_i)$ stands for the discretized distributions living on the computational lattice domain. The lattice nodes are linked by a set of velocity vectors $\cv_i$ (in the present case, we have $i=1,\ldots, 9$ in 2D and $i=1,\ldots,19$ in 3D). Furthermore, $f_i\ueq$ is a given equilibrium distribution and $f_i^F$ is a forcing-term that gives rise to a physical body force $\Fv$ in the Navier-Stokes equations (see eq.~\eqref{nse_nonideal} below). In the present case, this body force mediates the non-ideal fluid effects. For the specific forms of $f_i\ueq$ and $f_i^F$ as well as details of the implementation we refer to the original publication \cite{lee_eliminating_2006}.
The physical observables density $\rho$ and flow velocity $\uv$ can be obtained at any time-step from the lowest-order moments of $f$,
\begin{align} \rho &= \sum_i f_i = \sum_i f_i\ueq,\label{lb_den}\\
\rho\uv &= \sum_i f_i \cv_i  + \onehalf \Fv = \sum_i f_i\ueq \cv_i\,,
\label{lb_vel}
\end{align}

As a consequence of the LB dynamics of the present model, the density $\rho$ and velocity $\uv$ fulfil the Navier-Stokes equations of an isothermal, compressible non-ideal fluid, which consist of a continuity equation 
\beq \partial_t \rho + \nabla\cdot (\rho \uv) = 0\,,
\label{contin_eq}
\eeq
and a momentum equation,
\beq
 \partial_t(\rho \uv) + \nabla(\rho \uv\uv)= \Fv + \eta \nabla^2\uv + \left(\zeta +[1-2/d]\eta \right)\nabla\nabla\cdot\uv\,.
\label{nse_nonideal}
\eeq
Here, $\eta$ and $\zeta$ are the shear and bulk viscosity, which can be specified as input to the simulation. 
The body force $\Fv$ is responsible for the non-ideal character of the fluid and can be derived from a free energy functional. The thermodynamics of a non-ideal fluid is governed by a Ginzburg-Landau free-energy functional $\mathcal{F}$ of the form
\beq
\mathcal{F}[\rho] = \int d\rv\left[ \frac{\kappa}{2}|\nabla \rho|^2 + f_0(\rho) \right]\,,
\label{fef_rho}
\eeq
with $\kappa$ being a square-gradient parameter and $f_0$ a Landau free energy density.
Here, we take $f_0$ to be a simple double-well potential with freely prescribable minima around the equilibrium densities $\rho_V$, $\rho_L$ \cite{RowlinsonWidom_book, lee_eliminating_2006}
\beq  f_0(\rho) = \beta (\rho - \rho_V)^2(\rho-\rho_L)^2\,.
\label{f0}
\eeq
$\beta$ is a free parameter controlling the compressibility of the model.
From $\mathcal{F}$ we can define a generalized chemical potential
\beq \mu = \frac{\delta \mathcal{F}}{\delta \rho}=\partial_\rho f_0 -\kappa \nabla^2\rho
\label{chempot}
\eeq 
and finally the driving force that enters the Navier-Stokes eq.~\eqref{nse_nonideal}
\beq \Fv=-\rho \nabla \mu.
\eeq
In equilibrium, the free energy functional eq.~\eqref{fef_rho} admits for coexistence of liquid and vapor phases separated by a diffuse interface \cite{anderson_diffuse_1998}. In the case of a flat interface located at position $x=0$, the density profile is given by 
\beq \rho\st{int}(x) = \onehalf (\rho_L+\rho_V) + \onehalf (\rho_L-\rho_V) \tanh\left(\frac{x}{w}\right)\,,
\label{intprof}
\eeq
where 
\beq w=\frac{2}{\rho_L-\rho_V}\sqrt{\frac{2\kappa}{\beta}}\,
\eeq
is a measure for the interface width.

\subsection{Setup}

\begin{table*}[tb]
\begin{center}
\begin{tabular}{c | c c c c c c c | c c c c c c}
& \multicolumn{7}{|c|}{2D} & \multicolumn{6}{c}{3D}\\
\hline
Label & $\eta_L/\eta_V$ & $\sigma$ & $\tau_v$ & $\tau_i$ & \multicolumn{2}{c}{Oh}  & Re & $\eta_L/\eta_V$ & $\sigma$ & $\tau_v$ & $\tau_i$ & Oh & Re \\
    &                 & ($\times 10^{-4}$ l.u.)  & \multicolumn{2}{c}{($\times 10^4$ l.u.)} & l & s &          &               & ($10^{-4}$ l.u.) & \multicolumn{2}{c}{($\times 10^4$ l.u.)} & \\
\hline\hline
$\circ$ &              10 &    4.3  & 44 & 40 & 0.33 & 1.2 &   0.13 &           10 &   4.1 & 16 & 4.8 &  3.3 &  0.01 \\
$\vartriangle$  &     1000 &  6.6   & 30 & 31 & 0.27 & 0.97 &  0.22 &            100 &   8.0 & 9.2 & 3.4 &  2.6 &  0.03\\
{\tiny$\square$} &     100 &   1.3  & 16 & 23 & 0.20 & 0.70 &  0.40 &          1000 &    16 & 4.8 & 2.5 &  1.9 &    0.05 \\
$\bullet$ &            10 &    3.3  & 80  & 45 & 0.52 & 1.8 &  0.05 &             10 &    3.3 & 22 & 5.7 & 4.0 &  0.006 \\
$\blacktriangle$ &     100 &   13   & 26 & 23 & 0.33 & 1.2 &   0.16 &              4 &    13 & 5.2 & 2.7 &  2.0 &  0.03 \\
\hline
\end{tabular}
\end{center}
\caption{Parameters for our simulations in three and two dimensions. The first column refers to the symbols used in the plots, $\eta_L/\eta_V$ the liquid-vapor viscosity ratio (which is equal to the liquid-vapor density ratio), $\sigma$ the surface tension (in l.u.), $\tau_v$ and $\tau_i$ are the viscous and inertial times (in l.u.), Oh is the Ohnesorge number [eq.~\eqref{ohnesorge}] and Re is the average Reynolds number (computed by averaging the maximum velocity in the system over the full time domain of the coalescence process). In the 2D case, the letters `l' and `s' below Oh refer to the simulations performed using a large ($R_0=5000$) or small ($R_0=400$) droplet radius. The values for $\tau_v$ and $\tau_i$ are stated for the smaller radius.
}
\label{tab:parameters}
\end{table*}

In the 2D case, we study two resting droplets of identical radii of $R_0=5000$ lattice units (l.u.), initially separated by 4 l.u., that are placed into a periodic rectangular domain of size $20100\times 10200$ l.u.
For comparison, we also performed simulations using droplet radii of $R_0=400$ and box sizes of $1800\times 1000$ l.u., but keeping all the other simulation parameters the same as in the case of large radius. The discussion of the velocity field in sec.~\ref{sec:gen_asp} is based on simulations of droplets with radii of $R_0=1600$.
Due to computational limitations, in the 3D case simulations generally have system sizes of $460\times 250 \times 250$, where two spherical droplets each of radius $R_0=100$ l.u. and initially separated by 4 l.u. are placed.
The initial separation of the two droplets has been varied in a few cases, showing that this parameter has a negligible influence on our results.
Also, in a number of cases, simulations with larger system sizes but identical droplet radii have been performed to ensure that periodic boundary conditions have negligible influence on the evolution of the bridge radius. 
The bridge radius $b$ is determined from the (interpolated) position along the horizontal symmetry axis of the simulation box that corresponds to a density value of $(\rho_L+\rho_V)/2$. 
We follow the time evolution until the bridge radius has reached roughly half of the radius $R_0$ of the original droplets.
We find that the volume of the liquid droplet stays constant during the whole coalescence process with a relative change of the order of $10^{-4}$.
We have chosen comparable material parameters (i.e., viscosity, surface tension and density contrast between inner and outer fluid) to the silicon oil used in \cite{aarts_coalescence_2005}, amounting to typical droplet radii of a few mm and time units between 0.1 and 2~$\mu$s per l.u.. The viscous and inertial time scales are given by
\beq \tau_v = \frac{\eta R_0}{\sigma}\,,
\label{t-visc}
\eeq
and
\beq \tau_i = \sqrt{\frac{\rho_L R_0^3}{\sigma}}\,,\label{t-inert}
\eeq
respectively, and range between 20 and 100~ms when expressed in physical units.
Detailed simulation parameters can be found in Table \ref{tab:parameters}.
We cover a wide range of values for surface tension, viscosity ratio (which is equal to the density ratio here), Ohnesorge and Reynolds number. The Ohnesorge number is given by
\beq \text{Oh} = \frac{\tau_v}{\tau_i} = \frac{\eta}{\sqrt{\rho_L \sigma R_0}} = \sqrt{\frac{b}{R_0 \mathrm{Re}}}
\label{ohnesorge}
\eeq 
and has been recently shown to be an important quantity separating different coalescence regimes \cite{paulsen_pnas2012}. According to the phase-diagram proposed in \cite{paulsen_pnas2012}, the present values of the Ohnesorge number indicate that our simulations are located in the inertially-limited viscous regime. We will return to this point in secs.~\ref{sec:res}B and ~\ref{sec:summary}.

As our LB simulations are based on a diffuse interface model \cite{anderson_diffuse_1998, lee_eliminating_2006}, some care is needed in order to avoid spurious influences of the finite interface width on the determination of the bridge radius.
Denoting by $\xi$ the characteristic length scale over which the liquid-vapor interface is diffuse in our simulations, we have to make sure that we only consider the regime where all our physically interesting quantities are significantly larger than $\xi$, thereby approximating the sharp interface limit. This leads to the restrictions $b/\xi \gg 1$ for the bridge radius and $r_\sigma/\xi \gtrsim 1$ for the radius of curvature, $r_\sigma$, of the neck (see Fig.~\ref{fig:droplets}).
While the first condition is easily fulfilled in our simulations, the latter one imposes a lower limit on the physically meaningful bridge radius. From the geometric considerations of section \ref{sec:theory}, we have [eq.~\eqref{curv-rad}] $r_\sigma = b^2/2R_0$, and thus arrive at the condition
\beq \frac{b}{R_0} \gtrsim \sqrt{\frac{2 \xi }{ R_0}} \,.\label{b-cond} \eeq
With a typical value of $\xi=4$, we obtain $b/R_0\gtrsim 0.04$ for $R_0=5000$, $b/R_0\gtrsim 0.15$ for $R_0=400$ and $b/R_0\gtrsim 0.3$ for $R_0=100$.

%-------------------------------------------------------------------------------------------------------------

\section{Results}
\label{sec:res}

\subsection{General aspects}
\label{sec:gen_asp}

\begin{figure*}[t]
    (a)\includegraphics[width=0.47\linewidth]{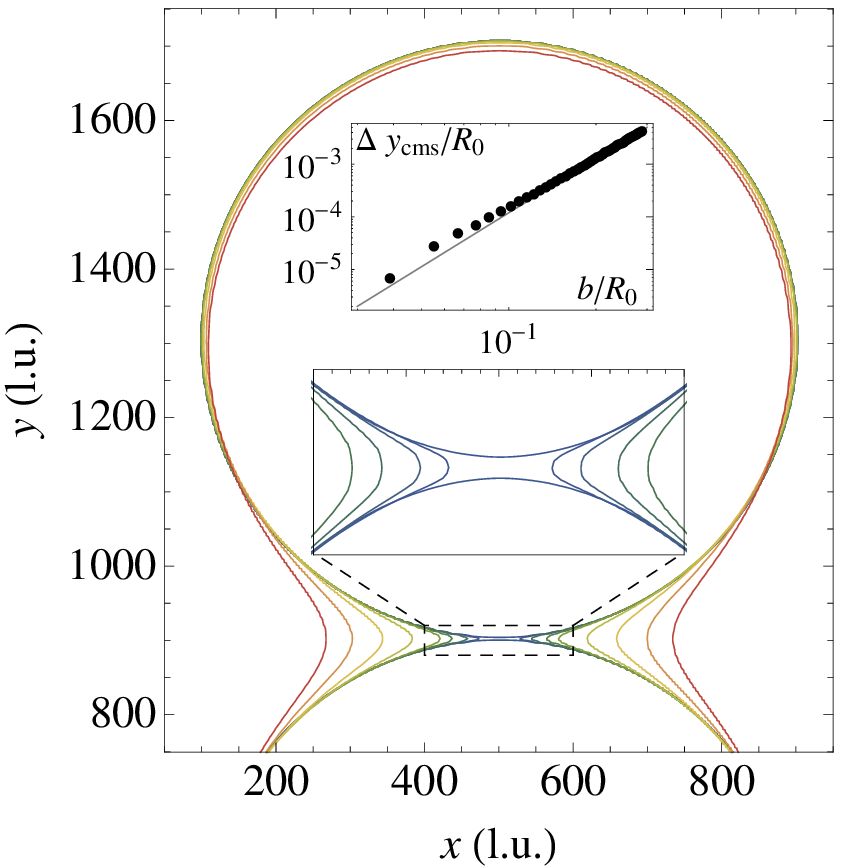}\qquad
    (b)\includegraphics[width=0.43\linewidth]{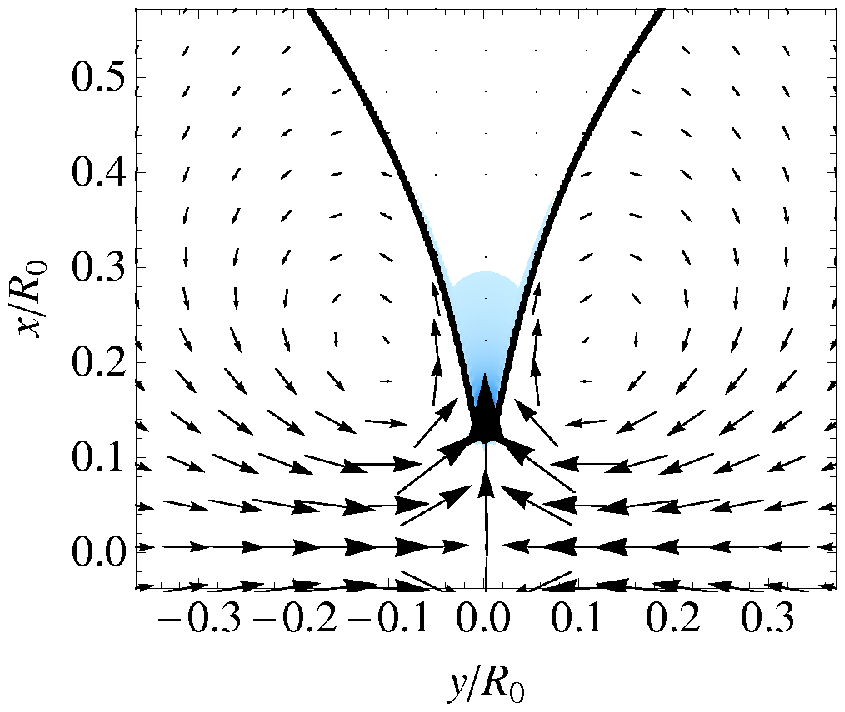}
   \caption{(a) Coalescence of two droplets initially at rest (in 2D). Contours represent the liquid-vapor interface and correspond to times 0, 0.01, 0.025, 0.05, 0.075, 0.15, 0.25, 0.375, 0.5 (in units of $\tau_v$, from inner to outer). The bottom inset is a magnification of the bridge region. The upper inset shows the center-of-mass position, $\Delta y\st{cms}\equiv y\st{cms}(t)-y\st{cms}(0)$ of one of the droplets (measured along the symmetry axis) in a double-logarithmic representation. The solid line represents the prediction of Hopper \cite{hopper_1984}. (b) Momentum field $\rho \uv$ at $t \simeq 0.075\tau_v$ in the region of the bridge for a typical simulation with a density ratio of $\rho_L/\rho_V=100$. The color-field represents the divergence of the velocity field, $\nabla\cdot \uv$ (the darker the color the more negative the value; positive values are very small and not resolved by the color scaling, corresponding to white in the plot).}
    \label{fig:coalesc_gen}
\end{figure*}

In Fig.~\ref{fig:coalesc_gen}a, a typical coalescence event in 2D as observed in our simulations is visualized.
The upper inset of Fig.~\ref{fig:coalesc_gen}a shows the evolution of the center-of-mass position $\Delta y\st{cms}\equiv y\st{cms}(t)-y\st{cms}(0)$ of one of the droplets in dependence of the bridge radius $b/R_0$. For all our simulations, we find that the data is well described (including the proportionality constant) by a power-law $\Delta y\st{cms}/R_0\simeq 0.3 (b/R_0)^{3.4}$, represented by the solid line in the inset of Fig.~\ref{fig:coalesc_gen}a.

We remark that if one plots $\Delta y\st{cms}/R_0$ vs.\ $t/\tau_v$ rather than vs.\ $b/R_0$, a general power-law exponent of $1.7$ is found, effectively in agreement with the predictions of the Stokesian theory of coalescence by Hopper \cite{hopper_1984}.
In this representation, however, the prefactor turns out to be strongly dependent on the droplet size. The exponent $3.4$ that describes the dependence of $\Delta y\st{cms}$ on $b/R_0$ can be obtained from Hopper's theory by substituting the proper time-dependence of the bridge radius as observed in the present simulations ($b\sim t^{1/2}$, see below).
We also note that, in contrast to predictions of \cite{eggers_coalescence_1999}, we have never observed the presence of a vapor bubble around the neck.

Fig.~\ref{fig:coalesc_gen}b shows the momentum field, $\rho\uv$, in the region close to the bridge typically observed in our simulations for a large density ratio (here, $\rho_L/\rho_V=100$). 
We observe a strong mass transport from the interior of each drop towards the bridge, driven by the difference in Laplace pressure between the drop and negatively curved neck. 

Interestingly (see color field in Fig.~\ref{fig:coalesc_gen}b), we find that the velocity field has a negative divergence ($\nabla\cdot\uv<0$) in the neck region, indicative of condensation (cf.~\cite{condensation}).
The possibility of phase transition is in fact a characteristic feature of the present simulation model. The occurrence of condensation can be understood as a consequence of the Kelvin-(Gibbs-Thomson-) effect \cite{landau_statmech_book} and the fact that we study coalescence of liquid droplets that are initially in equilibrium with their vapor phase: 
while the droplets are separated, the interface curvature is positive throughout, implying a slightly elevated chemical potential corresponding to an increased saturation vapor density (reckoned with respect to a flat interface) \cite{note_vap}. In contrast, the strongly negatively curved neck established after contact entails a local depression of the chemical potential, which in turn drives a condensation flux from the outer vapor regions towards the neck.

\begin{figure*}[tb]
	(a)\includegraphics[width=0.4\linewidth]{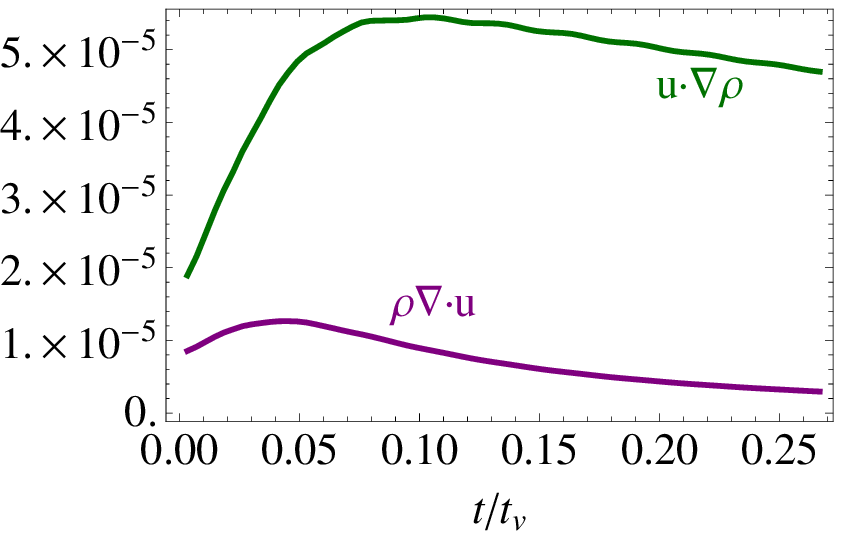}\qquad
	(b)\includegraphics[width=0.4\linewidth]{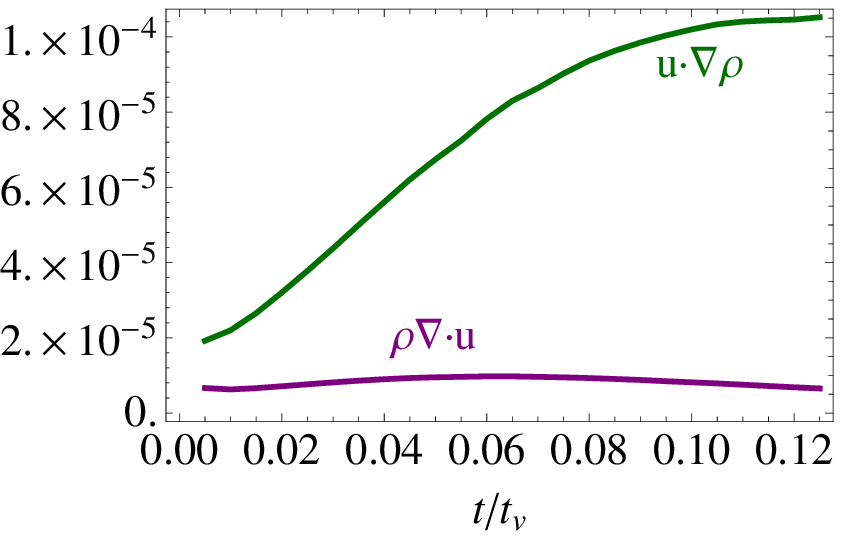}
   \caption{Comparison of the different mechanisms of mass transport [see eq.~\eqref{eq:cont}] contributing to the interface motion near the neck, for a density ratio of (a) $\rho_L/\rho_V=10$ and (b) $\rho_L/\rho_V=1000$. Shown are the contributions from the advective term $\uv\cdot\nabla\rho$ and the divergence term $\rho\nabla\cdot\uv$, computed as a weighted average over the interface near the neck.}
    \label{fig:veldiv}
\end{figure*}

For a more detailed study of this issue, we compare in Fig.~\ref{fig:veldiv} the different transport mechanism that contribute to the motion of the interface and their time dependence. The continuity equation [eq.~\eqref{contin_eq}], 
\beq \partial_t \rho = -\rho\nabla\cdot\uv -\uv\cdot\nabla\rho\,,
\label{eq:cont}\eeq
implies that the interface can move both due to advection, represented by $\uv\cdot\nabla\rho$, and evaporation/condensation, represented by $\rho\nabla\cdot \uv$. 
To compare the contribution of both terms, we compute them as a weighted average over the interface in a narrow region near the neck, i.e., $\int d\rv |\nabla\rho(\rv)| s(\rv) / \int d\rv |\nabla\rho| $, where $s(\rv)$ is either $\rho\nabla\cdot\uv$ or $\uv\cdot\nabla\rho$.
As seen from Fig.~\ref{fig:veldiv}, advection represents the dominant mechanism of interface motion in our system in all cases. As expected, the relative contribution form condensation increases at early time as the density ratio is reduced, since the compressibility of the model is enhanced. However, in the interesting late-time regime, we find that condensation is suppressed by almost an order of magnitude compared to advection, indicating that our model can be treated as effectively incompressible in this regime.

\subsection{Time-evolution of the bridge: 2D}
\label{sec:time-evol2d}
\begin{figure*}[tb]
   (a)\includegraphics[width=0.45\linewidth]{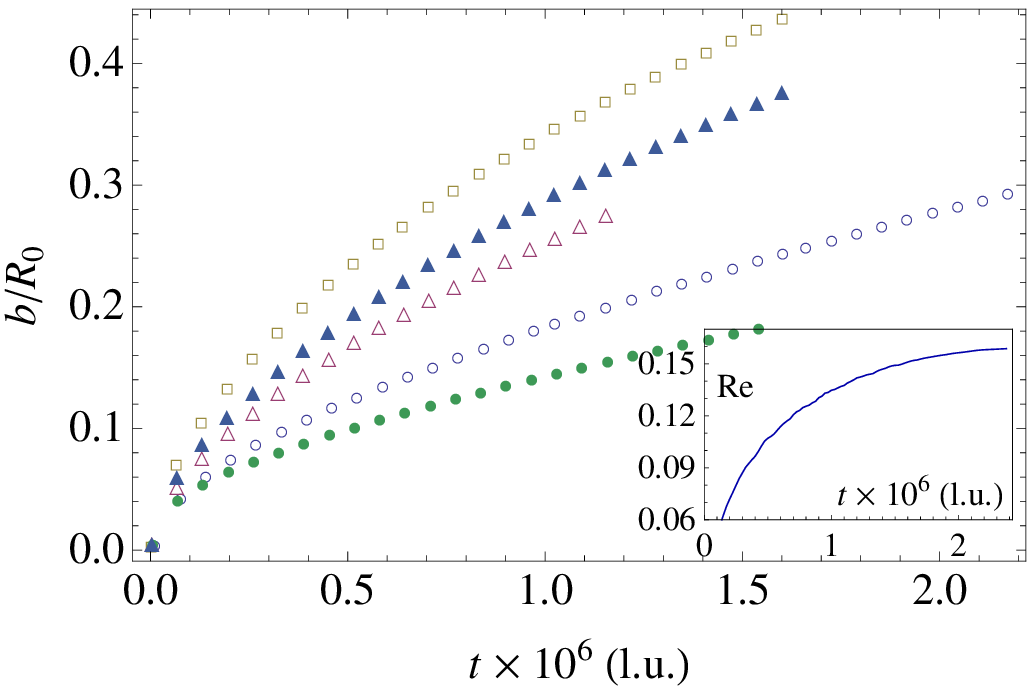}\qquad
   (b)\includegraphics[width=0.45\linewidth]{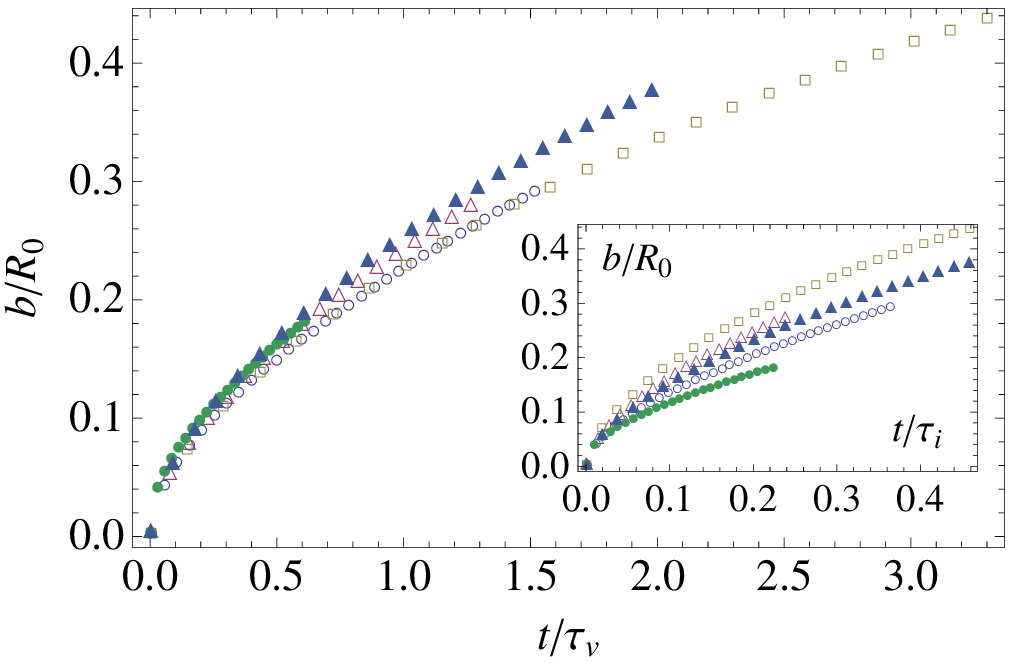}\\
   (c)\includegraphics[width=0.45\linewidth]{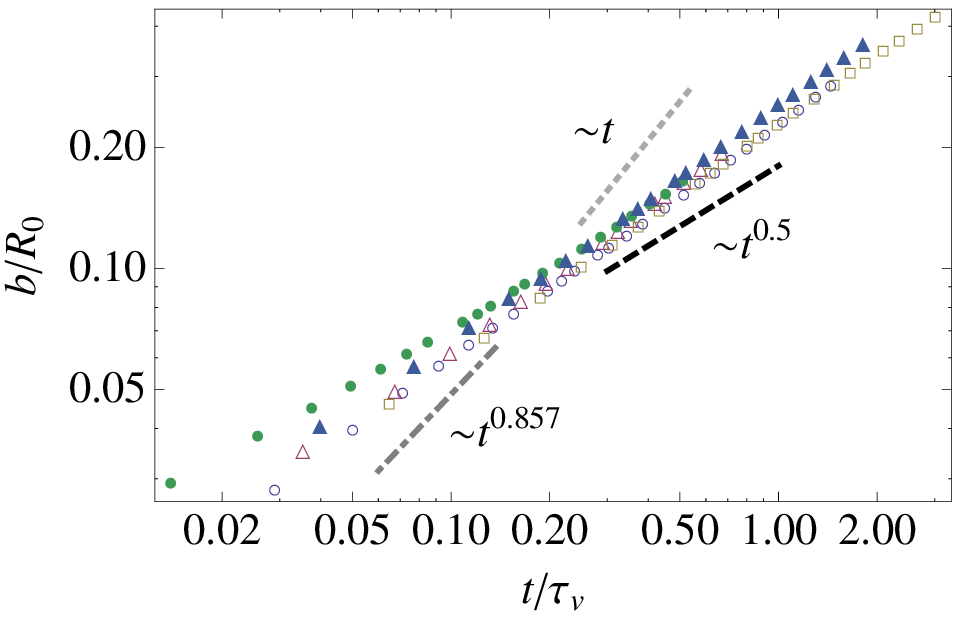}
   \caption{Viscous coalescence in 2D for two droplets of initial radius $R_0=5000$ l.u. (a) Raw data for the time evolution of the bridge radius $b$ (in units of the droplet radius $R_0$) as obtained from our simulations. The inset shows the evolution of the Reynolds number for one case. In (b), the data is shown when the time is expressed in terms of the viscous time $\tau_v$. For comparison, in the inset, time is rescaled by the inertial time $\tau_i$.  In (c), the data is plotted in a double-logarithmic representation using the same rescaling of the axes as in (b). The dashed, dot-dashed and dotted lines represent power laws $\sim t^{1/2}$, $\sim t^{0.857}$ and $\sim t$, respectively. See Table~\ref{tab:parameters} for a legend to the different symbols.}
    \label{fig:data2d}
\end{figure*}

\begin{figure*}[tb]
    (a)\includegraphics[width=0.45\linewidth]{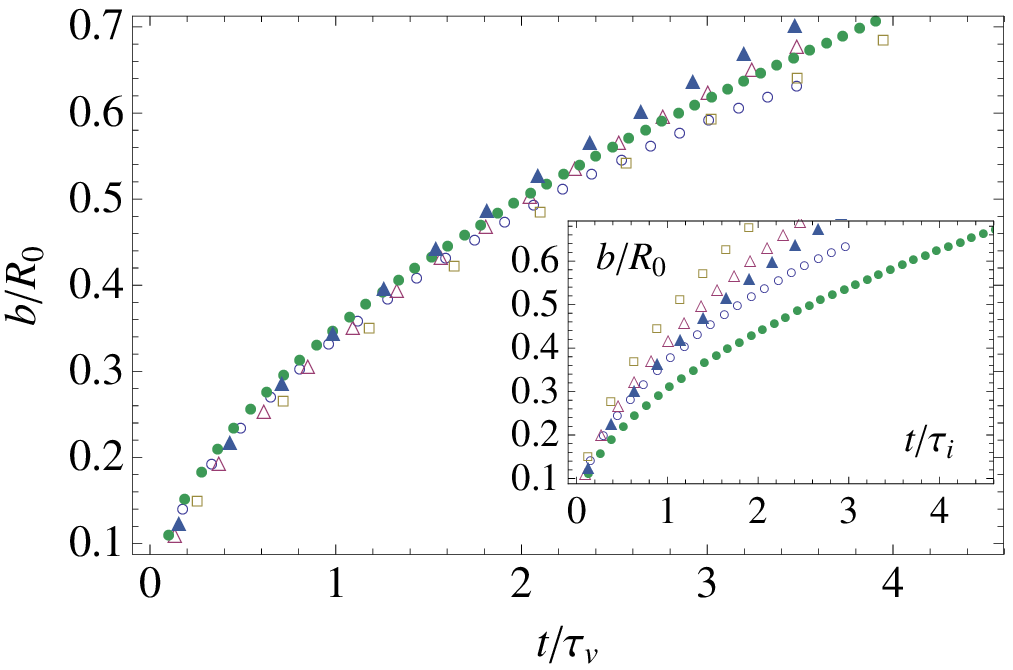}\qquad
    (b)\includegraphics[width=0.45\linewidth]{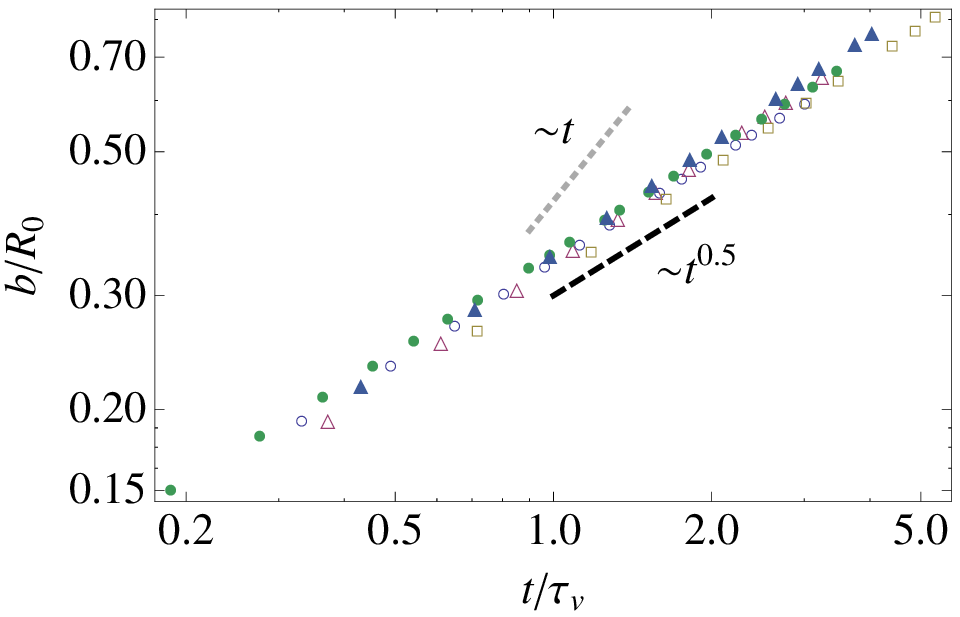}\\
   \caption{Viscous coalescence in 2D for two droplets of initial radius $R_0=400$ l.u. Time is expressed in terms of the viscous time $\tau_v$, while the bridge radius $b$ is expressed in units of $R_0$. For comparison, in the inset, the time is rescaled by the inertial time $\tau_i$.  In (a), the data is shown in a linear and in (b) in a double-logarithmic representation. The dashed and dotted lines represent power laws $\sim t^{1/2}$ and $\sim t$, respectively. See Table~\ref{tab:parameters} for a legend to the different symbols.}
    \label{fig:data2d_short}
\end{figure*}

We now turn to a detailed investigation of the time evolution of the bridge radius in our simulations.
Fig.~\ref{fig:data2d}a contains the raw data of the time evolution of the bridge radius obtained from five different simulations in two dimensions (for simulation parameters see Table \ref{tab:parameters}).
From the values of the Reynolds number (see inset) it can already be inferred that the coalescence process is located in the viscous regime \cite{note_reynolds}. This is also corroborated by the (approximate) data collapse obtained when rescaling the time axis with the viscous time $\tau_v$ [eq.~\eqref{t-visc}], as seen in Fig.~\ref{fig:data2d}b. In contrast, when expressing the time in units of the inertial time $\tau_i$, no collapse is obtained.
As the double-logarithmic representation of Fig.~\ref{fig:data2d}c shows, the bridge radius approximately follows a power-law $b(t)\sim (t/\tau_v)^{a}$ with an exponent $a$ varying between $0.50$ and $0.58$ within the individual curves. Note that there are no fitting parameters involved in obtaining these plots.
Interestingly, this behavior is consistent with the prediction obtained from a simple scaling argument based on the incompressible Stokes equations [see section \ref{sec:theory}].

For comparison, Fig.~\ref{fig:data2d_short} shows the evolution of the bridge radius obtained for droplets of radius $R_0=400$, using the same set of simulation parameters as in Fig.~\ref{fig:data2d}. Due to the smaller droplet radii, the evolution spans only about one decade in time. Nevertheless, by rescaling the time by the viscous time scale (Fig.~\ref{fig:data2d_short}a) and plotting the data in logarithmic representation (Fig.~\ref{fig:data2d_short}b) we again conclude that the droplets coalesce in the viscosity-dominated regime, following a $t^{1/2}$-law in time. 

\begin{figure*}[tb]
   \includegraphics[width=0.45\linewidth]{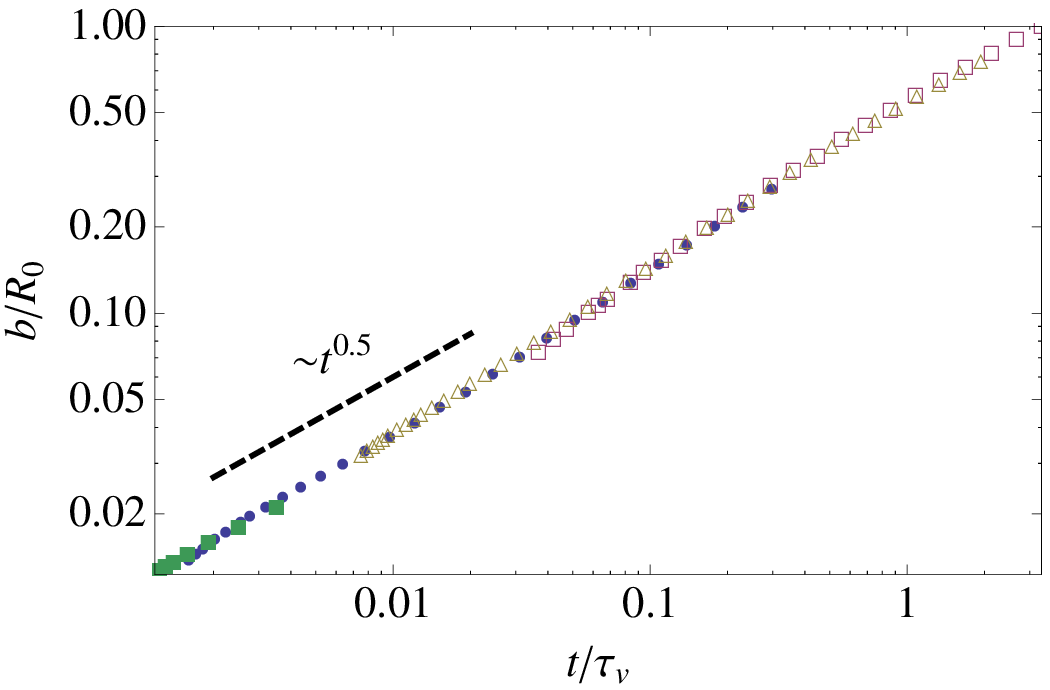}
   \caption[]{Effect of grid refinement on the scaling behavior of the bridge radius $b$ in 2D. For a fixed interface thickness of $\xi=4$ (l.u.), we vary the drop radius while keeping the Ohnesorge number constant. The droplet radii are chosen as $R_0=400$ ({\tiny $\square$}), $1600$ ($\vartriangle$), $5000$ ($\bullet$), $10000$ ({\tiny $\blacksquare$})  (all in l.u.). This corresponds to increasingly sharper interfaces: $\xi/R_0=0.02$ ({\tiny $\square$}), $0.005$ ($\vartriangle$), $0.0016$ ($\bullet$), $0.0008$ ({\tiny $\blacksquare$}). Time is expressed in terms of the viscous time scale $\tau_v$. In all cases, the density ratio is $\rho_L/\rho_V=1000$, the Ohnesorge number is $0.27$ and the interface width as well as the initial separation of the droplets are kept constant at 4 lattice units.}
\label{fig:grid_refinement}
\end{figure*}
In order to examine the possible effect of grid resolution on our results, we have systematically increased the spatial resolutions at a constant Ohnesorge number [eq.~\eqref{ohnesorge}]. Since the interface thickness is specific to the numerical scheme and does not have a direct counterpart in the corresponding macroscopic hydrodynamic equations, we have kept in these studies the number of interface grid points constant so that a higher resolution also corresponds to a sharper interface. Results on the time evolution of the bridge radius obtained from these simulations are shown in Fig.~\ref{fig:grid_refinement}. In plotting the data, we neglect the short initial transient which does not obey eq.~\eqref{b-cond}. As seen from Fig.~\ref{fig:grid_refinement}, upon rescaling $b$ by $R_0$ and the time $t$ by the viscous time $\tau_v$, all curves settle on a master curve $\sim t^{0.5}$. We thus conclude that, already for the smallest investigated droplet sizes of $R_0=400$, the underlying grid resolution is sufficiently high in order to resolve the correct hydrodynamic behavior.

We remark that experiments on viscous coalescence of two-dimensional liquid lenses \cite{burton_prl2007} reported a linear time evolution for the bridge radius (similar to the 3D case), rather than a $t^{1/2}$-scaling law.
Based on a recent study of Paulsen et al.~\cite{paulsen_pnas2012}, one would expect our simulations to reside in the inertially-limited viscous regime, which is empirically characterized by a linear time evolution. 
For comparison, an analytical theory for Stokesian coalescence by Hopper \cite{hopper_1984} predicts that $b\propto (t/\tau_v)^{0.857}$ in the range $0.0042 < b/R_0 < 0.21$ (note that this is an approximation to the full solution provided by Hopper \cite{hopper_1984}, valid during the whole coalescence process).
However, it is clearly visible from Fig.~\ref{fig:data2d}b and c that there appears no sufficiently extended region where either a linear or $t^{0.857}$-law can describe the data well.
We will return to a discussion on possible reasons for these discrepancies in section \ref{sec:summary}.

\subsection{Time-evolution of the bridge: 3D}

\label{sec:res_3d}

\begin{figure*}[tb]
   (a)\includegraphics[width=0.45\linewidth]{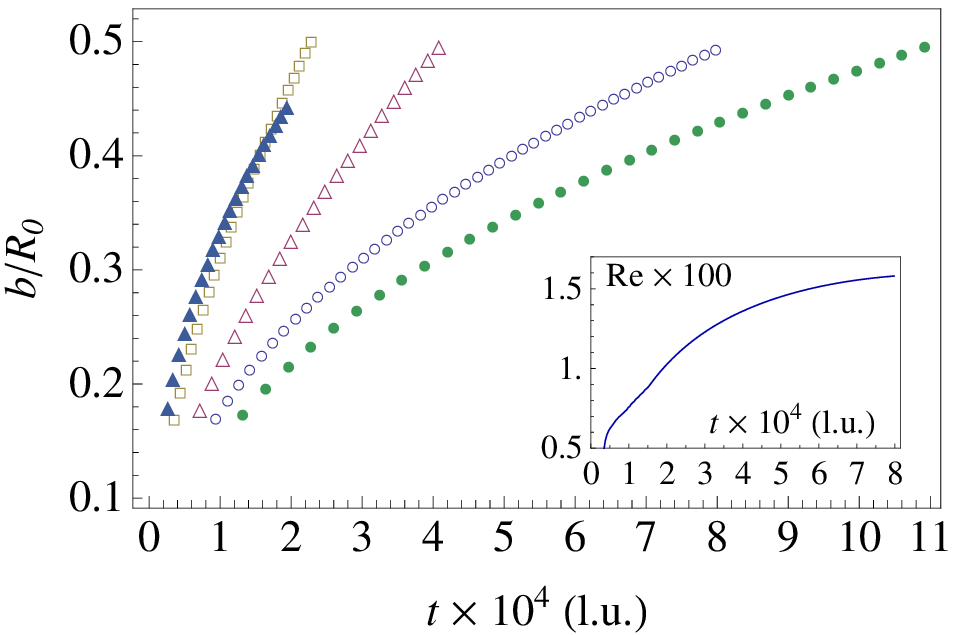}\qquad
   (b)\includegraphics[width=0.45\linewidth]{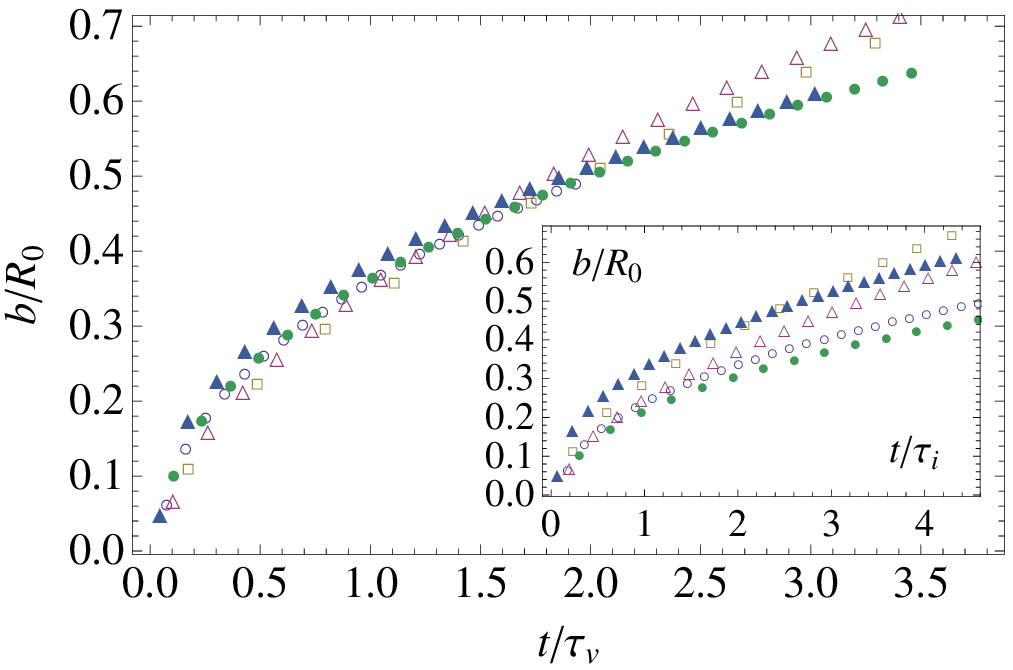}\\
   (c)\includegraphics[width=0.45\linewidth]{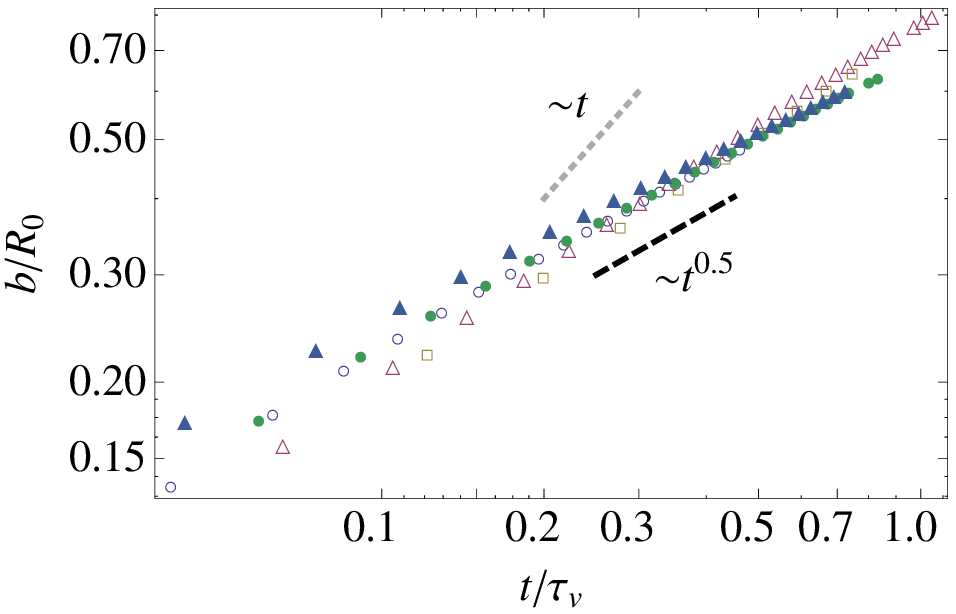}
   \caption{Viscous coalescence in 3D. (a) Raw data for the time evolution of the bridge radius $b$ (in units of the droplet radius $R_0$) as obtained from our simulations. The inset shows the evolution of the Reynolds number for one case. In (b), the data is shown when the time is expressed in terms of the viscous time $\tau_v$. For comparison, in the inset, time is rescaled by the inertial time $\tau_i$.  In (c), the data is plotted in a double-logarithmic representation using the same scaling of the axes as in (b). The dashed and dotted lines represent power laws $\sim t^{1/2}$ and $\sim t$, respectively. See Table~\ref{tab:parameters} for a legend to the different symbols.}
    \label{fig:data3d}
\end{figure*}

Turning to coalescence in the three dimensions, we analyze our data in an analogous manner as in the 2D case.
In Fig.~\ref{fig:data3d}a we show the bridge radius evolution obtained from five representative simulation runs (see Tab.~\ref{tab:parameters} for simulation parameters). Although simulation parameters are comparable to 2D, the Reynolds number remains roughly one order of magnitude smaller than in two dimensions. 
As seen in Fig.~\ref{fig:data3d}b, expressing the time in units of the viscous time, $\tau_v$, produces a reasonable scaling collapse of all data points onto a master curve. In contrast, no collapse is found when rescaling the time with the inertial time $\tau_i$ (inset to Fig.~\ref{fig:data3d}b). This confirms, independently from the value of the Reynolds number, the dominance of viscosity in the coalescence process.
As the double-logarithmic representation of Fig.~\ref{fig:data3d}c shows, the data follows a power law $b(t)\sim (t/\tau_v)^{1/2}$ with reasonable accuracy. The power-law exponent is found to be only approximately equal to $1/2$, varying in the range $0.45-0.6$ between the individual curves. 
Again, no fitting parameters are involved in the presentation of the data.
Note that, due to computational limitations, the initial radii of the two droplets had to be chosen significantly smaller than in 2D and, consequently, the time axis in Fig.~\ref{fig:data3d} spans only roughly one decade. This range is nevertheless comparable to experiments \cite{aarts_coalescence_2005, yao_viscous_2005}.

\begin{figure*}[t]
    (a)\includegraphics[width=0.45\linewidth]{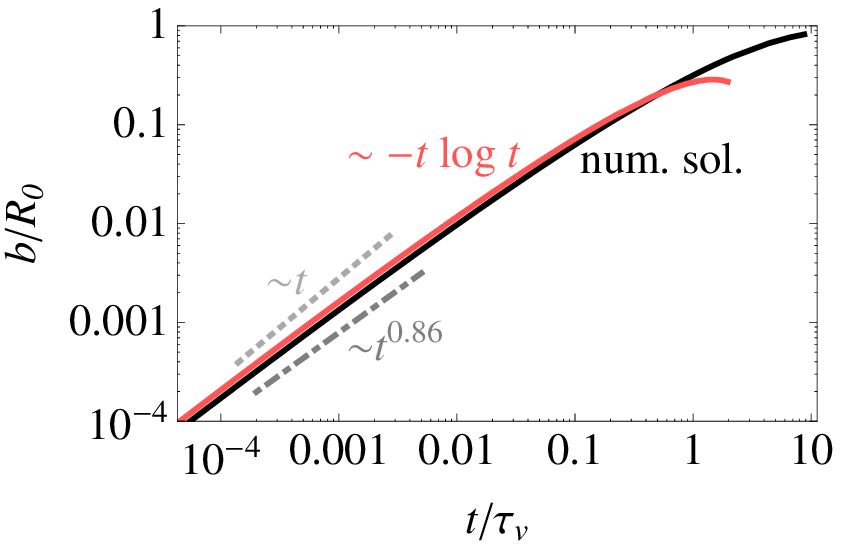}\qquad
    (b)\includegraphics[width=0.45\linewidth]{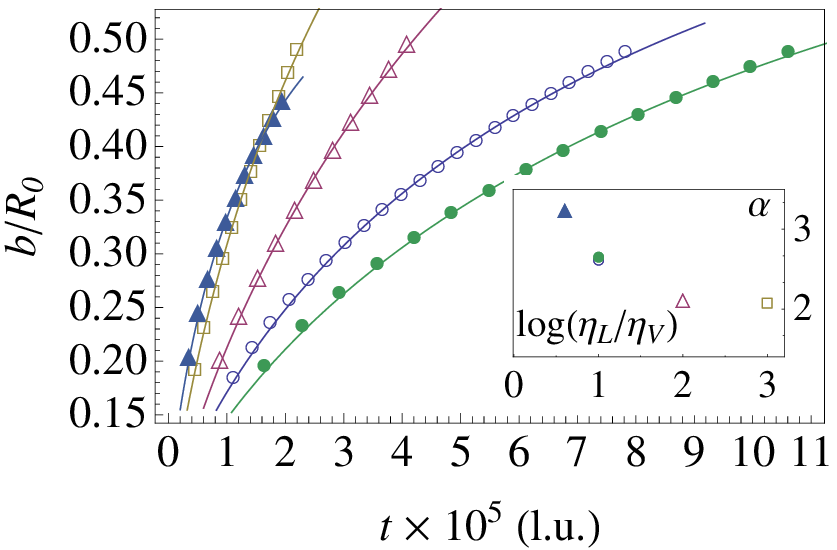}
   \caption{(a) Theoretical time evolution of the bridge radius according to the numerical solution of eq.~\eqref{eggers-deq} (thin solid curve) and its asymptotic approximation [eq.~\eqref{tlogt}, thick solid curve], for a typical set of droplet parameters. At early times, the analytical predictions can be well approximated by an effective power-law $b\sim t^{0.86}$ (dot-dashed line). For comparison, also a linear power-law ($b\sim t$, dotted) is indicated. (b) Numerical solution of eq.~\eqref{eggers-deq} (solid curves) fitted to the data for the bridge radius $b$ obtained from simulations in 3D (symbols). The inset shows that, in contrast to theoretical expectation \cite{eggers_coalescence_1999}, the fit parameter $\alpha$ decreases with increasing liquid-vapor viscosity ratio. Error bars are of the order of the symbol size.}
    \label{fig:eggers-fits3d}
\end{figure*}

Similarly to the two-dimensional case, even when acknowledging the slight variation in the power-law exponent, our results stand in contrast to experiments \cite{aarts_coalescence_2005, yao_viscous_2005, paulsen_prl2011}, where typically a linear time-dependence is observed (see, however, \cite{thoroddsen_remark}).
It is also interesting to compare our results to the analytical theory of three-dimensional coalescence of Eggers et al.~\cite{eggers_coalescence_1999}, which predicts the bridge radius (in the region $r_b\lesssim b \lesssim R_0$) to evolve according to the differential equation
\beq
\frac{\dot b(t/\tau_v)}{R_0} = -\frac{1}{2\pi} \log\left(c_0\, \bigg[\frac{b(t/\tau_v)} {R_0}\bigg]^{\alpha-1}\right)\,,
\label{eggers-deq}
\eeq
Here, $c_0$ and $\alpha$ are constants that, in the original work \cite{eggers_coalescence_1999}, are predicted as $c_0=1$ and $\alpha=3$ for an inviscid outer fluid while $\alpha=3/2$ for finite viscosity ratios (both valid for $b/R_0\ll 1$). For $b\lesssim 0.03 R_0$, the solution of eq.~\eqref{eggers-deq} follows an approximate scaling law of the form \cite{eggers_coalescence_1999}
\beq \frac{b\left(t/\tau_v\right)}{R_0} \sim \frac{t}{\tau_v} \log\left( \frac{t}{\tau_v}\right)\,.
\label{tlogt}
\eeq
In Fig.~\ref{fig:eggers-fits3d}a, the typical behavior of the numerical solution of eq.~\eqref{eggers-deq} as well as the approximate law \eqref{tlogt} is shown.
Several interesting observations can be made:
First of all, over many decades, the bridge radius as predicted by the analytical theory evolves significantly slower than it would be for a linear time-dependence \cite{note_logt}.
In fact, in the considered range, the behavior of eqs.~\eqref{eggers-deq} and \eqref{tlogt} can be well approximated by an effective power-law $\sim t^{n}$, with an exponent $n$ that is close to the value obtained by Hopper in the two-dimensional case \cite{hopper_1984}, $n\approx 0.86$ (the value of $n$ increases for still earlier decades in time).
At later times, the evolution of the bridge slows down and a time window appears where $b$ approximately evolves proportionally to $ t^{0.5}$.

Given the latter observation, it is tempting to fit the numerical solution of eq.~\eqref{eggers-deq} to our data for the bridge radii (Fig.~\ref{fig:eggers-fits3d}b). We find that reasonable agreement can only be obtained if $\alpha$ and $c_0$ are treated as fit parameters. In the inset to Fig.~\ref{fig:eggers-fits3d}b it is seen that the obtained best-fit values for the scaling exponent $\alpha$ decrease with the liquid-vapor viscosity ratio $\eta_L/\eta_V$, settling around a value of 2 for large viscosity ratios -- a trend which is opposite to the theoretical predictions of \cite{eggers_coalescence_1999}.
The parameter $c_0$ is found to be of the order of unity for all except the smallest viscosity ratios -- a result which roughly agrees with the theoretical expectation, $c_0=1$.
We remark that, due to the limited range in time, it can presently not be decided whether our observation of a $t^{1/2}$-law in the data for three-dimensional coalescence (Fig.~\ref{fig:data3d}c) is a manifestation of a finite-size effect associated with the slowing-down of the coalescence process or whether it actually represents a true deviation from the typical time evolution in the Stokesian theory of coalescence \cite{eggers_coalescence_1999}, similar to the two-dimensional case. Here, additional simulations over a much larger range of time will be necessary to clarify this issue. This is reserved for future work.

\section{Scaling considerations}
\label{sec:theory}
In sec.~\ref{sec:gen_asp} we observed that the contribution of condensation to the motion of the bridge is much reduced compared to advection and, in particular, has no significant influence on the time-evolution of the bridge radius. Indeed, we found the scaling law $b\sim (t/\tau_v)^{1/2}$ to be quite robust over a wide region of the parameter space, with the exponent generally being in the range $0.5\pm 0.1$.
Motivated by this findings, we present here a simple derivation of this scaling law based on the incompressible Stokes equations -- that is, neglecting evaporation or condensation from the outset. 

\begin{figure}[b]
    \includegraphics[width=0.52\linewidth]{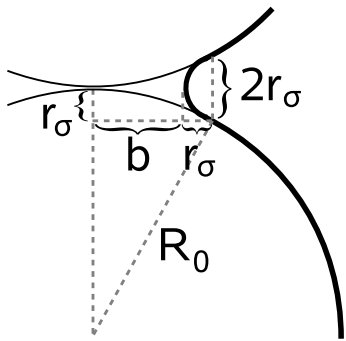}
   \caption{Geometry of the bridge region of two coalescing droplets as considered in our scaling theory. $R_0$ is the droplet radius, $b$ the radius of the connecting bridge and $r_\sigma$ the radius of curvature of the meniscus.}
    \label{fig:sketch}
\end{figure}

Fig.~\ref{fig:sketch} shows the basic situation we consider in our derivation. $R_0$ is the initial radius of each of the two droplets, $b$ the radius of the connecting bridge, and $\rsig$ the radius of curvature of the meniscus.
Due to rotational symmetry, it suffices to consider the problem in a plane that contains the conjoining line of the centers of the two droplets.
By geometry, $R_0^2 = (b+\rsig)^2 + (R_0-\rsig)^2$, and thus, neglecting $2r^2_\sigma$, we obtain the radius of curvature as
\beq \rsig = \frac{b^2}{2 R_0}\cdot \frac{1}{1-\frac{b}{R_0}} \approx \frac{b^2}{2 R_0}\,,
\label{curv-rad}
\eeq
where the last approximation is justified since we focus on $b\ll R_0$. It should be remarked that a similar result is also used in \cite{paulsen_prl2011}, but a number of previous works \cite{hopper_1984, eggers_coalescence_1999} found $r_\sigma \sim b^3$.
To obtain a relation for the time dependence of the bridge radius, we apply scaling arguments similar in spirit to \cite{eggers_coalescence_1999, duchemin_coalescence_2003, narhe_epl2008} to the incompressible Navier-Stokes equations \cite{Landau_FluidMech},
\beq
\rho (\partial_t + \bv{u} \cdot \nabla)\bv{u} = -\nabla p + \eta \nabla^2 \bv{u}\,,
\label{eq:NS}
\eeq
where $\rho$ is the density, $\uv$ the fluid velocity and $p$ the local pressure.
We shall neglect a possible rigid translation of the coalescing droplets (although, in principle, this effect can be relevant \cite{hopper_1984, paulsen_pnas2012}) and also ignore the influence of the vapor on the dynamics.
Focusing now on the axis along the direction of the bridge radius, the flow velocity $u$ near the meniscus is, by continuity, determined by the motion of the bridge alone, $u = \dot b$.
Assuming that the pressure varies between zero at the center of the bridge and the Laplace pressure $p_L\simeq -\sigma / \rsig$ at the curved meniscus, we can approximate $\nabla p\sim p_L/b$.
Finally, we shall assume that the velocity varies smoothly (e.g., parabolically) between the center of the bridge (where $u\approx 0$) and the interface, allowing us to estimate $\nabla^2 u \sim -u/b^2$.

In the \emph{viscous} regime, eq.~\eqref{eq:NS} becomes $\eta \nabla^2 u = \nabla p_L$ in the steady-state and above simplifications lead to
\beq
\dot b = \frac{2 R_0 \sigma}{\eta} \frac{1}{b}\,.
\label{visc-law}
\eeq
The solution of this differential equation is easily written down as
\beq
b(t)^2-b_0^2 = R_0^2\frac{t-t_0}{\tau}\,,
\label{scaling-sol}
\eeq
where $t_0$ is a suitably chosen initial time, $b_0\equiv b(t_0)$ and the characteristic time scale $\tau$ is (up to a numerical prefactor) given by the viscous time $\tau_v/4 = \eta R_0/4\sigma$ [eq.~\eqref{t-visc}].
For completeness, we state also the result for the \emph{inertial} regime, where the advection term $\rho(\bv{u}\cdot\nabla)\bv{u}$ dominates over the viscous stress. Here, the same solution as in eq.~\eqref{scaling-sol} is obtained, but with $\tau$ replaced by the inertial time $\tau_i/\sqrt{8} = \sqrt{\rho R_0^3/8\sigma}$ [eq.~\eqref{t-inert}].
It must be emphasized here that the above results hold independent of the spatial dimension.

Neglecting the integration constants $t_0$ and $b(t_0)$ in \eqref{scaling-sol}, which is justified except in the early stages of the evolution, the growth-law for the bridge radius can be expressed as
\beq \frac{b(t)}{R_0} \sim \left(\frac{t}{\tau_{v,i}}\right)^{1/2}\,.
\label{sqrt-law}
\eeq
In the inertial regime, the above relation is well known \cite{duchemin_coalescence_2003} and has been confirmed by experiments \cite{inertial_coal, aarts_coalescence_2005, yao_viscous_2005, burton_prl2007, paulsen_prl2011}.
Remarkably, our simple scaling theory yields a power-law with the same exponent $1/2$ also in the viscous regime, albeit with a different characteristic time scale.
We note that relation \eqref{sqrt-law} agrees with Frenkel's classic result on viscous sintering \cite{frenkel_1949}, where, however, it was obtained under the assumption that the two coalescing droplets reduce their surface free energy by moving closer together as a whole, without considering explicitly the dynamics of the bridge.

Of course, the scaling arguments leading to eq.~\eqref{sqrt-law} are not rigorous and therefore have to be taken with care. For instance, approximating the Laplacian of the velocity at the bridge as $\nabla^2 u\sim u / b^{n}$ with an arbitrary power $n$ allows one to write down a generalization of eq.~\eqref{visc-law} as
\beqn
\dot b = \frac{2 \sigma}{\eta} \left(\frac{b}{R_0}\right)^{n-3}\,.
\eeqn
Obviously, the choice $n=2$ corresponds to eq.~\eqref{sqrt-law}, supported by our simulations, while $n=4$ would lead to a linear time dependence of $b$, in agreement with experiments.

\begin{figure*}[tb]
   (a)\includegraphics[width=0.45\linewidth]{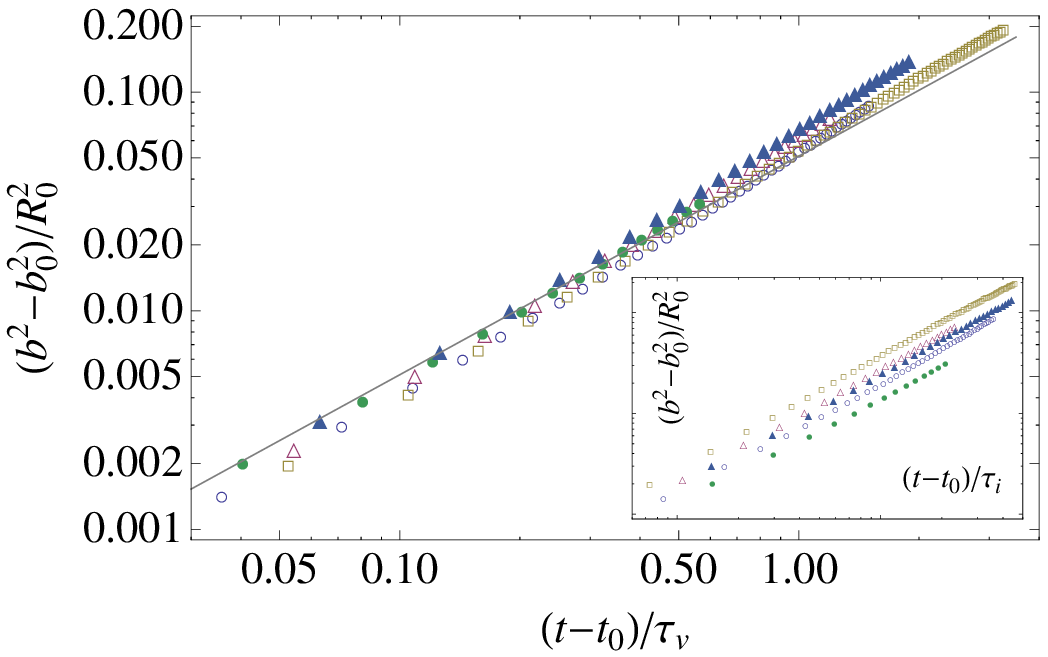}\qquad
   (b)\includegraphics[width=0.45\linewidth]{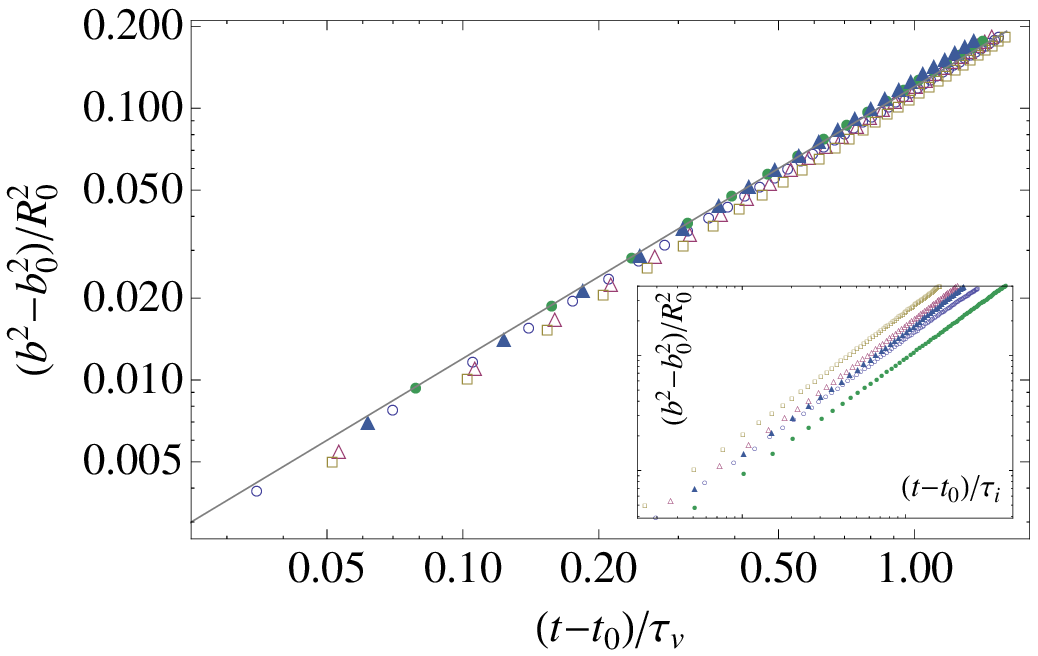}\\
   (c)\includegraphics[width=0.45\linewidth]{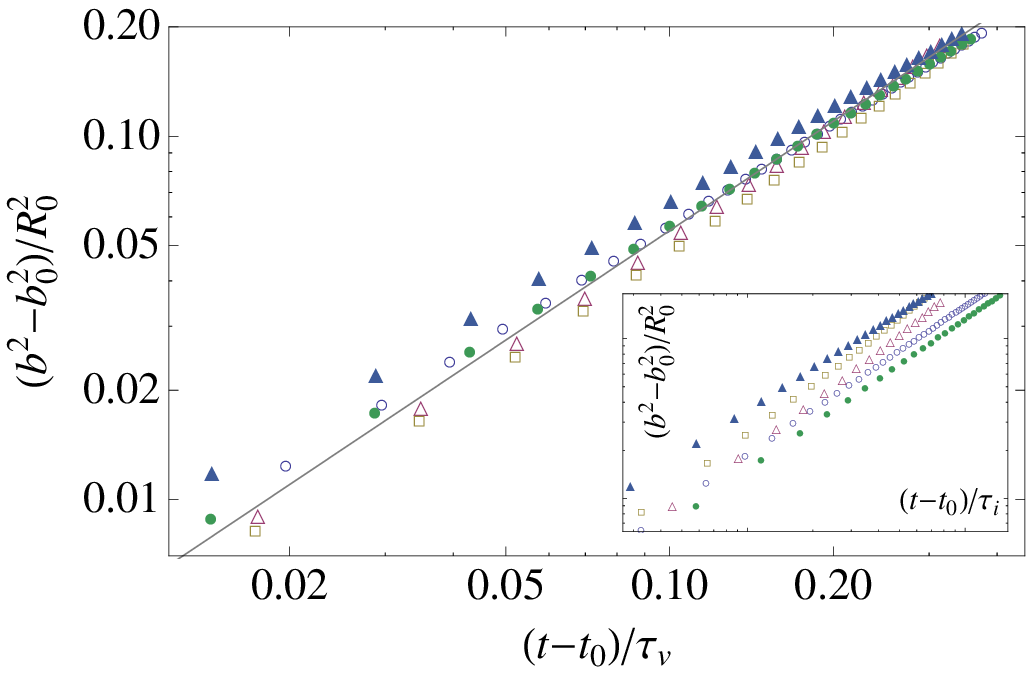}
   \caption[]{Viscous coalescence of two identical droplets in (a,b) 2D and (c) 3D. 
In the main figures, the bridge radius data are transformed according to the relation \eqref{scaling-sol}, where $b_0=b(t_0)$ is fixed by the choice of $t_0$. 
Droplet radii are (a) $R_0=5000$, (b) $R_0=400$ and (c) $R_0=100$ l.u. 
The inset illustrates the lack of scaling collapse if the time is expressed in units of the inertial time.}
\label{fig:alt_scal}
\end{figure*}
In sec.~\ref{sec:res} we have shown that the time evolution of the bridge radius observed in our simulations is consistent with a scaling relation of the form of eq.~\eqref{sqrt-law}, without invoking any fit parameters.
Assuming for the moment the validity of the present derivation, we can make use of the freedom to choose the initial value of the time $t_0$ as allowed by the solution \eqref{scaling-sol} of our scaling ansatz. The bridge radius data transformed according to the relation \eqref{scaling-sol},where $b_0=b(t_0)$ is fixed by the choice of $t_0$, is shown in Fig.~\ref{fig:alt_scal}a,b in the two-dimensional and in Fig.~\ref{fig:alt_scal}c in the three-dimensional case. Consistent with the results in sec.~\ref{sec:res}, we find reasonably good scaling collapse of all data points if time axis is scaled by the viscous time, and no collapse if the inertial time is used (see insets).

\section{Summary}
\label{sec:summary}
In this work, we have investigated the coalescence of two identical resting droplets in the viscous regime in two and three dimensions.
From Lattice Boltzmann simulations of the full isothermal Navier-Stokes equations for a non-ideal fluid, we find that the time evolution of the bridge radius can be well described by a $t^{1/2}$-scaling law -- in striking similarity to what is generally found to hold in the inertial regime, albeit with a different characteristic time constant.
A negative divergence of the velocity field is observed at the neck of the coalescing droplets, indicating the presence of condensation. However, over the investigated parameter range, condensation is found to give a negligible contribution to the motion of the interface at the neck, the dominant process being advection. 
Guided by this observation, we have shown that our simulation results can be rationalized through simple scaling arguments applied to the incompressible Stokes equations [see eq.~\eqref{scaling-sol}].

Our findings differ markedly from recent experiments \cite{aarts_coalescence_2005, yao_viscous_2005, burton_prl2007,  paulsen_prl2011}, which report a linear time-evolution of the bridge radius in the viscous regime both in two and three dimensions.
In two dimensions, our findings are also clearly different from analytical theories \cite{hopper_1984}, which predict a power-law $b\sim t^{0.86}$.
However, the center-of-mass motion of each droplet is found to be described by a power-law $\sim t^{1.7}$, with an exponent that, surprisingly, agrees well with the theoretical prediction \cite{hopper_1984}. 
Due to the limited range in time, no definite conclusions concerning possible agreement with the theory for three-dimensional Stokesian coalescence \cite{eggers_coalescence_1999} can be drawn at present.
It is also useful to note that, in the framework of the phase-diagram of \cite{paulsen_pnas2012}, our simulations would not be located in the true Stokes regime, which applies to droplets of much larger viscosity than presently used, but instead in the inertially-limited viscous regime. However, this does not resolve the above mentioned discrepancies since most experiments also reside in this region \cite{aarts_coalescence_2005, burton_prl2007, paulsen_prl2011}. Furthermore, the time-evolution of the bridge radius in the inertially-limited viscous regime has only empirically been found to follow a linear time evolution, with a theoretical derivation of this result lacking so far.
Presently, we have no explanation for the discrepancies between our work and existing theories or experiments. This is a subject of future work.

\begin{acknowledgements}
We thank K.\ Stratford and D.\ Bonn for useful discussions and an anonymous referee for valuable suggestions. Financial support by the Deutsche Forschungsgemeinschaft (DFG) under the grant number Va205/3-3 (within the Priority Program SPP1164) is acknowledged. ICAMS acknowledges funding from its industrial sponsors, the state of North-Rhine Westphalia and the European Commission in the framework of the European Regional Development Fund (ERDF).
\end{acknowledgements}

%\appendix*


\begin{thebibliography}{}
\bibitem{solgel-science}C. J. Brinker and G. W. Scherrer, \textit{Sol-Gel Science} (Academic Press, London, 1990).
\bibitem{nikolayev_prl1996}V. S. Nikolayev, D. Beysens and P. Guenoun, ``New hydrodynamic mechanism for drop coarsening,'' Phys. Rev. Lett. {\bf 76}, 3144 (1996).
\bibitem{raindrops_1995}Z. Hu and R. C. Srivastava, ``Evolution of raindrop size distribution by coalescence, breakup, and evaporation: Theory and observations,'' J. Atmos. Sci. {\bf 52}, 1761 (1995).
\bibitem{paulsen_prl2011}J. D. Paulsen, J. C. Burton and S. R. Nagel, ``Viscous to inertial crossover in liquid drop coalescence,'' Phys. Rev. Lett. {\bf 106} 114501, (2011).
\bibitem{aarts_coalescence_2005}D. G. A. L. Aarts, H. N. W. Lekkerkerker, H. Guo, G. H. Wegdam and D. Bonn, ``Hydrodynamics of droplet coalescence,'' Phys. Rev. Lett. {\bf 95} 164503, (2005).
\bibitem{duchemin_coalescence_2003}L. Duchemin, J. Eggers and C. Josserand, ``Inviscid coalescence of drops,'' J. Fluid Mech. {\bf 487}, 167 (2003).
\bibitem{burton_prl2007}J. C. Burton and P. Taborek, ``Role of dimensionality and axisymmetry in fluid pinch-off and coalescence,'' Phys. Rev. Lett. {\bf 98}, 224502 (2007).
\bibitem{lee_eliminating_2006}T. Lee and P. F. Fischer, ``Eliminating parasitic currents in the lattice Boltzmann equation method for nonideal gases,'' Phys. Rev. E {\bf 74}, 046709 (2006).
\bibitem{inertial_coal}A. Menchaca-Rocha, A. Martinez-Davalos, R. Nunez, S. Popinet and S. Zaleski, ``Coalescence of liquid drops by surface tension,'' Phys. Rev. E {\bf 63}, 046309 (2001);
M. Wu, T. Cubaud and C.-M. Ho, ``Scaling law in liquid drop coalescence driven by surface tension,'' Phys. Fluids {\bf 16}, L51 (2004);
S. C. Case and S. R. Nagel, ``Coalescence in low-viscosity liquids,'' Phys. Rev. Lett. {\bf 100}, 084503 (2008).
\bibitem{thoroddsen_2005}S. T. Thoroddsen, K. Takehara and T. G. Etoh, ``The coalescence speed of a pendent and a sessile drop,'' J. Fluid Mech. {\bf 527}, 85 (2005).
\bibitem{frenkel_1949}J. Frenkel, ``Viscous flow of crystalline bodies under the action of surface tension,'' J. Phys. (Moscow) {\bf 9}, 385 (1949).
\bibitem{sintering_exp}G. C. Kuczynski, ``Study of the sintering of glass,'' J. Appl. Phys. {\bf 20}, 1160 (1949);
W. D. Kingery and M. Berg, ``Study of the initial stages of sintering solids by viscous flow, evaporation-condensation, and self-diffusion,'' J. Appl. Phys. {\bf 26}, 1205 (1955).
\bibitem{hopper_1984}R. W. Hopper, ``Coalescence of Two Equal Cylinders: Exact Results for Creeping Viscous Plane Flow Driven by Capillarity'' J. Am. Ceram. Soc. {\bf 67}, C262 (1984).
\bibitem{eggers_coalescence_1999}J. Eggers, J. R. Lister and H. A. Stone, ``Coalescence of liquid drops,'' J. Fluid Mech. {\bf 401}, 293 (1999).
\bibitem{yao_viscous_2005}W. Yao, H. J. Maris, P. Pennington and G. M. Seidel, ``Coalescence of viscous liquid drops,'' Phys. Rev. E {\bf 71}, 016309 (2005). Here, an exponential scaling function with an almost linear regime at small times was found.
\bibitem{thoroddsen_remark}An exception to this is the experiment of \cite{thoroddsen_2005}, where, in a one case, an $t^{1/2}$-behavior was observed for $b\gtrsim 0.2 R_0$.
\bibitem{paulsen_pnas2012}J. D. Paulsen, J. C. Burton, S. R. Nagel, S. Appathurai, M. T. Harris and O. A. Basaran, ``The inexorable resistance of inertia determines the initial regime of drop coalescence'', Proc. Nat. Acad. Sci. {\bf 109}, 6857 (2012).
\bibitem{lbm_review}S. Succi, \textit{The Lattice Boltzmann Equation for Fluid Dynamics and Beyond} (Oxford University Press, Oxford, 2001); C. K. Aidun and J. F. Clausen, ``Lattice-Boltzmann method
for complex flows,'' Annu. Rev. Fluid Mech. {\bf 42} 439 (2010);
\bibitem{droplets_gen}P. Raiskinm\"{a}ki, A. Koponen, J. Merikoski and J. Timonen, ``Spreading dynamics of three-dimensional droplets by the lattice-Boltzmann method,'' Comput. Mater. Sci. {\bf 18}, 7 (2000); A. Dupuis and J. M. Yeomans, ``Modeling droplets on superhydrophobic surfaces: equilibrium states and transitions,'' Langmuir {\bf 21}, 2624 (2005); F. Varnik, P. Truman, B. Wu, P. Uhlmann, D. Raabe and M. Stamm, ``Wetting gradient induced separation of emulsions: A combined experimental and lattice Boltzmann computer simulation study,'' Phys. Fluids {\bf 20}, 072104 (2008); 
G. Falcucci, S. Chibbaro, S. Succi, X. Shan, and H. Chen, ``Lattice Boltzmann models for non-ideal fluids with arrested phase-separation,'' Phys. Rev. E {\bf 77}, 036705 (2008);
D. Chiappini, G. Bella, S. Succi and S. Ubertini, ``Applications of finite-difference lattice Boltzmann method to breakup and coalescence in multiphase flows,'' Int. J. Mod. Phys. C {\bf 20}, 1803 (2009);
M. Gross, F. Varnik, D. Raabe and I. Steinbach, ``Small droplets on superhydrophobic substrates,'' Phys. Rev. E {\bf 81}, 051606 (2010).
F. Varnik, M. Gross, N. Moradi, G. Zikos, P. Uhlmann, P. Mueller-Buschbaum, D. Magerl, D. Raabe, I. Steinbach and M. Stamm, ``Stability and dynamics of droplets on patterned substrates: insights from experiments and lattice Boltzmann simulations'', J. Phys.: Condens. Matter {\bf 23}, 184112 (2011).
\bibitem{kikkinides_2008}E. S. Kikkinides, A. G. Yiotis, M. E. Kainourgiakis and A. K. Stubos, ``Thermodynamic consistency of liquid-gas lattice Boltzmann methods: Interfacial property issues,'' Phys. Rev. E {\bf 78}, 036702 (2008).
\bibitem{gross_shearstress_pre2011}M. Gross, N. Moradi, G. Zikos and F. Varnik, ``Shear stress in nonideal fluid lattice Boltzmann simulations'', Phys. Rev. E {\bf 83}, 017701 (2011).
\bibitem{RowlinsonWidom_book}J. S. Rowlinson and B. Widom, \textit{Molecular Theory of Capillarity} (Oxford University Press, Oxford, 1989).
\bibitem{anderson_diffuse_1998}D. M. Anderson, G. B. McFadden and A. A. Wheeler, ``Diffuse-interface methods in fluids mechanics,'' Annu. Rev. Fluid. Mech. {\bf 30}, 139 (1998).
\bibitem{condensation}A. J. Briant, A. J. Wagner and J. M. Yeomans, ``Lattice Boltzmann simulations of contact line motion. I. Liquid-gas systems,'' Phys. Rev. E {\bf 69}, 031602 (2004); X. Xu, T. Qian, ``Contact line motion in confined liquid-gas systems: Slip versus phase transition,'' J. Chem. Phys. {\bf 133}, 204704 (2010).
\bibitem{landau_statmech_book}L. D. Landau and E. M. Lifshitz, \textit{Statistical Physics. Part I} (Pergamon, New York, 1980).
\bibitem{note_vap}Since the droplet radii are quite large in the present case, the increase of the vapor density is negligible and hence we can assume $\rho_V\ut{sat}\approx \rho_V$, where $\rho_V$ denotes the equilibrium vapor density over a flat interface.
\bibitem{note_reynolds}It is noteworthy that, recently \cite{paulsen_prl2011}, it has been suggested that the relevant characteristic length for the variations of the flow in the coalescence phenomenon is not the bridge radius, $b$, but rather the (half) width of the meniscus,  $r_\sigma\approx b^2/R_0$, which is essentially equal to the radius of curvature. This leads to a Reynolds number which is by a factor of $b/R_0$ smaller than the estimate given in Table \ref{tab:parameters}.
\bibitem{note_logt}The function $-t\log t$ approaches a linear power-law for $t\rightarrow 0$. We find empirically that a linear approximation becomes good only for $t / \tau_v < 10^{-15}$ on a logarithmic scale.
\bibitem{narhe_epl2008}R. D. Narhe, D. A. Beysens and Y. Pomeau, ``Dynamic drying in the early-stage coalescence of droplets sitting on a plate,'' EPL {\bf 81}, 46002 (2008).
\bibitem{Landau_FluidMech}L. D. Landau and E. M. Lifshitz, \textit{Fluid Mechanics} (Pergamon, New York, 1959).

\end{thebibliography}
\end{document}